\documentstyle[pra,epsf,aps]{revtex}
\begin{document}
\setlength{\oddsidemargin}{-1cm}
\setlength{\evensidemargin}{-1cm}
\newcommand{\nc}{\newcommand}
\nc{\beq}{\begin{equation}}
\nc{\eeq}{\end {equation}}
\nc{\bea}{\begin {eqnarray}}
\nc{\eea}{\end{eqnarray}}
\nc{\ba}{\begin{array}}
\nc{\ea}{\end{array}}
\nc{\nn}{\nonumber}
\nc{\bpi}{\begin{picture}}
\nc{\epi}{\end{picture}}
\nc{\scs}{\scriptstyle}
\nc{\sss}{\scriptscriptstyle}
\nc{\ts}{\textstyle}
\nc{\ds}{\displaystyle}
\nc{\mb}[1]{\mbox{#1}}
\nc{\unit}{{\mb{\boldmath\large $1$}}}
\nc{\half}{{\ts\frac{1}{2}}}
\nc{\jb}{\bar{J}}
\nc{\jh}{\hat{J}}
\nc{\od}{{\cal O}}
\nc{\p}{\partial}
\nc{\Ga}{\Gamma}
\nc{\al}{\alpha}
\nc{\be}{\beta}
\nc{\ga}{\gamma}
\nc{\ka}{\kappa}
\nc{\qq}{{\bf q}_k^2}
\nc{\pp}{{\bf p}^2}
\title{Efficient Algorithm for Perturbative Calculation of
Multiloop Feynman Integrals}
\author{Boris Kastening$^1$\thanks{Email: boris@thphys.uni-heidelberg.de}
and Hagen Kleinert$^2$\thanks{Email: kleinert@physik.fu-berlin.de,
%\hfil
%\protect\newline
URL: http://www.physik.fu-berlin.de/\~{}kleinert \hfil}}
\address{$^1$Institut f\"ur Theoretische Physik, Universit\"at Heidelberg,
Philosophenweg 16, 69120 Heidelberg, Germany\\
$^2$Institut f\"ur Theoretische Physik, Freie Universit\"at Berlin,
Arnimallee 14, 14195 Berlin, Germany}
%\date{}
\maketitle
\begin{abstract}
We present an efficient algorithm for
calculating  multiloop Feynman integrals perturbatively.
\end{abstract}
%\pacs{03.20.+i\\ 04.20.Fy\\ 02.40.+m}
%\pacs{}
%%%%%%%%%%%%
%
{}~\\
\noindent
{\bf 1.}
Recently, a new method has been proposed
to calculate Feynman integrals of multiloop
diagrams perturbatively \cite{K1}.
Together with the solution procedure for
graphical recursion relations developed in Ref.~\cite{rec},
this should ultimately lead to the completely automatized
computer generation of
perturbation expansions of field theories
up to high orders.
Such expansions
are needed in all
strongly coupled fluctuating field systems,
for example those
describing the critical phenomena close to second-order phase transitions
\cite{sc}.
So far, expansions have been limited to seven loops
only \cite{MN,seven}, which are barely sufficient to yield critical
exponents
\cite{seval}
with an accuracy comparable to experimental data \cite{Lipa}.

In this note, we would like to show how the expansions
proposed in \cite{K1} can be performed
most efficiently, such that they can be carried out on a computer to high
orders in a limited computer time.

{}~\\
\noindent
{\bf 2.}
A basic Feynman integral with $L$ loops,  $n$ internal lines, and
$E$ external momenta ${\bf k}_1,\ldots,{\bf k}_E$
has the form
\beq
I^D_{a_k}=\int\frac{d^Dp_1}{(2\pi)^D}\cdots\frac{d^Dp_L}{(2\pi)^D}
\prod_{k=1}^n\frac{1}{(1+{\bf q}_k^2)^{a_k}},  ~~~~n\ge L,
\label{@1}
\eeq
with some powers $a_k$, where
 ${\bf q}_k$
are the  momenta carried by the lines, and the integrations run over all
loop momenta ${\bf p}_i$.
The line momenta  ${\bf q}_k$
are linear combinations of the loop momenta  ${\bf p}_i$
and the external momenta ${\bf k}_j$.

For simplicity, we have set all masses equal to unity.
A Feynman integral
with different non-zero masses can be reduced to (\ref{@1})
by
an appropriate rescaling of the line momenta ${\bf q}_k$.
Following  Ref.~\cite{K1}, we view the integral (\ref{@1}) as a special
case $I^D_{a_k}=I^D_{a_k}(1)$ of the function
\beq
I^D_{a_k}(\ka)=\int\frac{d^Dp_1}{(2\pi)^D}\cdots\frac{d^Dp_L}{(2\pi)^D}
\prod_{k=1}^n\frac{e^{a_k(\ka-1){\bf q}_k^2}}{(1+\ka{\bf q}_k^2)^{a_k}}
\label{@inte}\eeq
to be calculated perturbatively
via a Taylor series expansions in powers of $ \kappa $.

It is the purpose of this note to point out
that the simplest way to derive such an expansion is by
rewriting each generalized propagator in the Schwinger
parametric form
\beq
\frac{e^{a(\ka-1){\bf q}^2}}{(1+\ka{\bf q}^2)^a}
=\frac{1}{\Ga(a)}\int_0^\infty dtt^{a-1}e^{-t}e^{[a(\ka-1)-\ka t]{\bf q}^2}.
\label{@3}\eeq
Then the integrals (\ref{@inte}) take the form
\bea
I^D_{a_k}(\ka)
&=&
\frac{1}{\Ga(a_1)}\cdots\frac{1}{\Ga(a_l)}
\int_0^\infty dt_1t_1^{a_1-1}e^{-t_1}\cdots
\int_0^\infty dt_nt_n^{a_n-1}e^{-t_n}
\nn\\
&&\times
\int\frac{d^Dp_1}{(2\pi)^D}\cdots\int\frac{d^Dp_L}{(2\pi)^D}
\exp\left\{-\sum_{k=1}^n[a_k(1{-}\ka)+\ka t_k]{\bf q}_k^2\right\}.
\eea
Collecting the $L$ loop momenta ${\bf p}_i$
and the $E$ external momenta ${\bf k}_j$ in
single vector symbols
\bea
p=({\bf p}_1,\ldots,{\bf p}_L),
\\
k=({\bf k}_1,\ldots,{\bf k}_E),
\eea
we rewrite
\bea
\sum_{k=1}^n[a_k(1{-}\ka)+\ka t_k]{\bf q}_k^2
&=&
\frac{1}{2}p^TMp+p^TM'k+\frac{1}{2}k^TM''k
\eea
and complete the squares to
\bea
\sum_{k=1}^n[a_k(1{-}\ka)+\ka t_k]{\bf q}_k^2
&=&
\frac{1}{2}\left(p+M^{-1}M'k\right)^TM\left(p+M^{-1}M'k\right)
+\frac{1}{2}k^T\left(M''-{M'}^TM^{-1}M'\right)k,
\eea
with symmetric matrices $M$ and $M''$.
%A summation over spatial indices is understood.
After a shift $p\rightarrow p-M^{-1}M'k$ of integration variables,
the ${\bf p}_k$ integrations become Gaussian, and we obtain
\bea
\label{iakdgeneral}
I^D_{a_k}(\ka)
&=&
\frac{1}{\Ga(a_1)\cdots\Ga(a_l)}
\int_0^\infty dt_1t_1^{a_1-1}e^{-t_1}\cdots
\int_0^\infty dt_lt_l^{a_l-1}e^{-t_l}
e^{-\frac{1}{2}k^T\left(M''-{M'}^TM^{-1}M'\right)k}
\int\frac{d^Dp_1}{(2\pi)^D}\cdots\int\frac{d^Dp_L}{(2\pi)^D}
e^{-\frac{1}{2}p^TMp}
\nn\\
&=&
\frac{(2\pi)^{-LD/2}}{\Ga(a_1)\cdots\Ga(a_l)}
\int_0^\infty dt_1t_1^{a_1-1}e^{-t_1}\cdots
\int_0^\infty dt_lt_l^{a_l-1}e^{-t_l}
\frac{
e^{-\frac{1}{2}k^T\left(M''-{M'}^TM^{-1}M'\right)k}}{(\det M)^{D/2}},
\label{@@9}\eea
where the matrices $M$, $M'$, and $M''$ depend on $\ka$ and the $t_k$
through linear combinations
\beq
c_k(\ka,t_k)\equiv a_k(1-\ka)+\ka t_k.
\eeq
Although the entries of the matrix $M$ depend on the routing of
the loop momenta through the different lines, the determinant of $ M$
is invariant under changes of the routing,
except for trivial relabelings of the $a_k$.

In order to derive the desired expansion of $I^D_{a_k}(\ka)$ in powers of
$\ka$,
we expand the integrand
on the right hand side of (\ref{iakdgeneral}) in powers of
$\ka$, whose coefficients are  polynomials in the parameters $t_i$,
$(i=1,\dots,L)$.
The $t_i$-integrals can then all be performed
using the formula
\beq
\label{trivint}
\int_0^\infty dtt^\ga e^{-t}=\Ga(\ga+1).
\eeq

For diagrams without external momenta, appearing in the perturbation
expansions for the ground state of quantum field theories,
(\ref{iakdgeneral}) simplifies to
\beq
\label{iakdvacuum}
I^D_{a_k}(\ka)
=\frac{(2\pi)^{-LD/2}}{\Ga(a_1)\cdots\Ga(a_l)}
\int_0^\infty dt_1t_1^{a_1-1}e^{-t_1}\cdots
\int_0^\infty dt_lt_l^{a_l-1}e^{-t_l}
(\det M)^{-D/2}.
\label{@@12}\eeq

More general Feynman integrals than those in Eq.~(\ref{@1})
may contain loop momenta ${\bf p}_k$ in
the numerator of the integrand. These can be calculated with a simple extension
of the above technique,
by introducing ``source terms" $\Sigma _{i=1}^{L}{\bf j}_i\cdot{\bf p}_i$
into the exponents of (\ref{@inte}) and (\ref{@3}), and appropriately
differentiate the resulting $\ka$-expansion with respect to ${\bf j}_i$,
which are set equal to zero at the end.

{}~\\
\noindent
{\bf 3.}
As a first example, take the exactly solvable one-loop integral
\beq
\label{@12}
I^D_{a}=\int\frac{d^Dp}{(2\pi)^D}\frac{1}{(1+\pp)^a}
=
\frac{\Ga(a-D/2)}{(4\pi)^{D/2}\Ga(a)}.
\label{@13}\eeq
Its $\ka$-generalized version can be expressed in terms of a confluent
hypergeometric function,
\bea
\label{oneloop}
\lefteqn{I_a^D(\ka)
\equiv
\int\frac{d^Dp}{(2\pi)^D}
\frac{e^{a(\ka-1)\pp}}{(1+\ka\pp)^a}
=
\frac{1}{(4\pi\ka)^{D/2}}
\Psi\left(\frac{D}{2},1+\frac{D}{2}-a;\frac{a(1-\ka)}{\ka}\right)}
\label{@14}\\
&=&
\frac{1}{(4\pi\ka)^{D/2}}\left[\frac{\Ga(a-D/2)}{\Ga(a)}
{}_1F_1\left(\frac{D}{2},1+\frac{D}{2}-a;\frac{a(1-\ka)}{\ka}\right)
+\frac{\Ga(D/2-a)}{\Ga(D/2)}(1-\ka)^{a-D/2}
{}_1F_1\left(a,1+a-\frac{D}{2};\frac{a(1-\ka)}{\ka}\right)\right]
\nonumber
\eea
with
\beq
{}_1F_1(\al;\be;z)
\equiv
\sum_{k=0}^\infty\frac{(\al)_k}{(\be)_k}\frac{z^k}{k!},
{}~~~~~~~~
(a)_s\equiv\frac{\Ga(a+s)}{\Ga(a)}=\prod_{r=0}^{s-1}(\al+r)
{}~~~({\rm Pochhammer's~symbol}).
\nonumber \eeq
In Ref.~\cite{K1} this was calculated perturbatively via a Wick expansion.
Here we use our general formula (\ref{@@12}) for vacuum integrals.
The number of loops is $L=1$, and we identify
\beq
q_1=p,
{}~~~a_1=a,
{}~~~c_1=a(1- \kappa )+ \kappa t,
{}~~~M=2(c_1),
{}~~~\det M=2c_1.
\eeq
Expanding
$(\det M)^{-D/2}$ in  powers of $ \kappa $,
and performing the resulting integrals over $t$
in Eq.~(\ref{@@12}),
we find  directly
the perturbation expansion for the loop integrals
(\ref{@13})
in any dimension $D$:
\bea
I_{a}^D(\ka)
&=&
\frac{1}{(4\pi a)^{D/2}}\bigg[
1 + {\frac{D\left( 2 + D \right)
      {\ka^2}}{8a}} -
  {\frac{D\left( 2 + D \right)
      \left( 4 + D \right) {\ka^3}}{24
      {a^2}}} + {\frac{\left( 2 + a \right)
      D\left( 2 + D \right)
      \left( 4 + D \right)
      \left( 6 + D \right) {\ka^4}}{128
      {a^3}}}
\nn\\
&&\hspace{11ex}
- {\frac{\left( 6 + 5a \right)
      D\left( 2 + D \right)
      \left( 4 + D \right)
      \left( 6 + D \right)
      \left( 8 + D \right) {\ka^5}}{960
      {a^4}}}
+\od(\ka^6)\bigg].
\eea
The expansion can easily extended any desired  order.
It agrees, of course,
with what we would obtain from the exact expression
(\ref{@14})
via a large-argument expansion
of the  confluent
hypergeometric function.

{}~\\
\noindent
{\bf 4.}
As a nontrivial example, take the integral of the watermelon diagram
treated in \cite{K1} only in $D=2$ dimensions:
\beq
I^D
=
\rule[-14pt]{0pt}{34pt}
\bpi(34,12)
\put(17,3){\circle{24}}
\put(17,3){\oval(24,8)}
\put(5,3){\circle*{4}}
\put(29,3){\circle*{4}}
\epi
=
\int\frac{d^Dp_1}{(2\pi)^D}\frac{d^Dp_2}{(2\pi)^D}\frac{d^Dp_3}{(2\pi)^D}
\frac{1}{1+{\bf p}_1^2}\frac{1}{1+{\bf p}_2^2}\frac{1}{1+{\bf p}_3^2}
\frac{1}{1+({\bf p}_1+{\bf p}_2+{\bf p}_3)^2}.
\eeq
This integral has the powers
\beq
a_1=a_2=a_3=a_4=1,
\eeq
and we identify the line momenta as
\beq
q_1=p_1,\;\;\;\;q_2=p_2,\;\;\;\;q_3=p_3,\;\;\;\;q_4=p_1+p_2+p_3,
\eeq
such that the matrix $M$ is
\beq
M=2\left(
\ba{ccc}
a_1+a_4&a_4&a_4\\
a_4&a_2+a_4&a_4\\
a_4&a_4&a_3+a_4
\ea
\right),
\eeq
\beq
\det M=8(a_1a_2a_3+a_1a_2a_4+a_1a_3a_4+a_2a_3a_4).
\eeq

For the function $I_{a_k}^D(\ka)$,
we then obtain in any dimension $D$
\bea
\label{idep}
\lefteqn{I^D(\ka)
=
\frac{1}{2^{4D}\pi^{3D/2}}\times}
\nn\\
&&
\bigg[
1 + {\frac{9D\left( 2 + D \right)
      {\ka^2}}{32}} -
  {\frac{9D\left( 2 + D \right)
      \left( 4 + D \right) {\ka^3}}{128}} +
  {\frac{3D\left( 2 + D \right)
      \left( 1048 + 522D + 81{{D}^2} \right)
        {\ka^4}}{4096}}
\nn\\
&&\hspace{2ex}
-
  {\frac{9D\left( 2 + D \right)
      \left( 4 + D \right)
      \left( 2576 + 918D + 117{{D}^2} \right)
        {\ka^5}}{40960}}
\nn\\
&&\hspace{2ex}
+
  {\frac{D\left( 2 + D \right)
      \left( 564864 + 397744D +
        110916{{D}^2} + 15228{{D}^3} +
        891{{D}^4} \right) {\ka^6}}{65536}}
\nn\\
&&\hspace{2ex}
-
  {\frac{3D\left( 2 + D \right)
      \left( 4 + D \right)
      \left( 29651840 + 15696528D +
        3452148{{D}^2} + 391068{{D}^3} +
        19683{{D}^4} \right) {\ka^7}}{9175040}}
\nn\\
&&\hspace{2ex}
   + {\frac{3D\left( 2 + D \right)
      \left( 4 + D \right)
      \left( 1419854080 + 843338336D +
        212508840{{D}^2} + 29562300{{D}^3} +
        2344950{{D}^4} + 85779{{D}^5} \right)
       {\ka^8}}{83886080}}
\nn\\
&&\hspace{2ex}
+\od(\ka^9)\bigg].
\eea
For $D=1$ this reduces to
\beq
I^1(\ka)=\frac{1}{2^4\pi^{3/2}}
\left[1 + {\frac{27{\ka^2}}{32}} -
  {\frac{135{\ka^3}}{128}} +
  {\frac{14859{\ka^4}}{4096}} -
  {\frac{97497{\ka^5}}{8192}} +
  {\frac{3268929{\ka^6}}{65536}} -
  {\frac{63271629{\ka^7}}{262144}} +
  {\frac{22569248565{\ka^8}}{16777216}}
+\od(\ka^{9})\right],
\eeq
and for $D=2$ to
\beq
I^2(\ka)
=
\frac{1}{2^8\pi^3}
\left[1 + {\frac{9{\ka^2}}{4}} -
  {\frac{27{\ka^3}}{8}} +
  {\frac{453{\ka^4}}{32}} -
  {\frac{1647{\ka^5}}{32}} +
  {\frac{15157{\ka^6}}{64}} -
  {\frac{157293{\ka^7}}{128}} +
  {\frac{3720699{\ka^8}}{512}}
+\od(\ka^9)\right],
\eeq
thus extending easily the expansions in \cite{K1}.

For $D=3$,
the expansion reads
\bea
I^3(\ka)
&=&
\frac{1}{2^{12}\pi^{9/2}}\bigg[
1 + {\frac{135{\ka^2}}{32}} -
  {\frac{945{\ka^3}}{128}} +
  {\frac{150435{\ka^4}}{4096}} -
  {\frac{1206387{\ka^5}}{8192}} +
  {\frac{48595005{\ka^6}}{65536}} -
  {\frac{1079675235{\ka^7}}{262144}}
\nn\\
&&\hspace{10ex}
+   {\frac{432899207685{\ka^8}}{16777216}}
+\od(\ka^9)\bigg].
\eea

{}~\\
\noindent
{\bf 5.}
Having developed the tools for
finding perturbation expansions of Feynman integrals,
it remains to study the large-order behavior, and to find
suitable  methods for the resummation of the expansions with high accuracy.
Together with the automatized generation of
the Feynman diagrams of Ref.~\cite{rec},
this will open the way for an ``industrial production"
of high-loop expansions for critical exponents.


\begin{thebibliography}{11}
\bibitem{K1}
H. Kleinert,
{\em Perturbative Calculation of Multi-Loop Feynman Diagrams.
New Type of Expansions for Critical Exponents\/},
Berlin Preprint 1999 (hep-th/9908078).


\bibitem{rec}

H. Kleinert, A. Pelster, B. Kastening, M. Bachmann,
{\em Recursive Graphical Construction of Feynman Diagrams
and Their Multiplicities in $\phi^4$- and $\phi^2\,A$-Theory\/},
Berlin Preprint 1999 (hep-th/9907168).

\bibitem{sc}
   H. Kleinert, Phys. Rev. D {\bf 57}, 2264 (1998) (E-Print aps1997jun25\_001);
addendum: Phys. Rev. D {\bf 58}, 107702 (1998) (cond-mat/9803268).



\bibitem{MN}D.B. Murray and B.G. Nickel, unpublished.


\bibitem{seven}
H. Kleinert,
Phys. Rev. D {\bf 60}, 085001 (1999)
(hep-th/9812197).

\bibitem{seval}
H. Kleinert
{\em Theory and Satellite Experiment
on Critical Exponent $ \alpha $ of Specific Heat in
Superfluid Helium\/},
FU-Berlin preprint 1999 (cond-mat/9906107).


\bibitem{Lipa}J.A. Lipa, D.R. Swanson, J. Nissen,
T.C.P. Chui and U.E. Israelson, {Phys. Rev. Lett.} {\bf 76}, 944 (1996).


\bibitem{nobeta}
   H. Kleinert,
         {\em Critical Exponents without beta-Function\/},
     FU-Berlin preprint 1999 (cond-mat/9906359)

\end{thebibliography}
\end{document}